\documentclass[preprint]{aastex631}
\usepackage{amsmath}
\usepackage[T1]{fontenc}

\begin{document}

\title{On the Dynamical Erasure of Initial Conditions in Multi-Planetary Systems}

\author[0009-0005-6067-0457]{Kevin Marimbu}
\affiliation{Department of Physics and Trottier Space Institute, McGill University, 3600 rue University, H3A 2T8 Montr\'eal QC, Canada}
\affiliation{Trottier Institute for Research on Exoplanets (iREx), Universit\'e de Montr\'eal, Montr\'eal, QC, Canada}

\author[0000-0002-1228-9820]{Eve J. Lee}
\affiliation{Department of Physics and Trottier Space Institute, McGill University, 3600 rue University, H3A 2T8 Montr\'eal QC, Canada}
\affiliation{Trottier Institute for Research on Exoplanets (iREx), Universit\'e de Montr\'eal, Montr\'eal, QC, Canada}

\begin{abstract}

Do sub-Neptunes assemble close to where we see them or do they form full-fledged farther away
from their host star then migrate inwards? We explore this question using the distribution of measured orbital periods, one of the most fundamental observable parameters. Under disk-induced migration, planet occurrence rate is expected to decrease towards shorter orbital periods. Presently, the observed sub-Neptune period distribution is flat in log period, between 10 and 300 days. We show, using N-body integration, how post-disk dynamical instabilities and mergers in multi-planetary systems erase the initial conditions of migration emplaced in period distributions over 10s to 100 Myr timescale, in rough agreement with an observational hint of the abundance of resonant pairs for systems younger than 100 Myr which drops dramatically for more evolved systems. We comment on caveats and future work.

\end{abstract}

\section*{}

Most mature multi-planetary systems are far from mean motion resonance (MMR) \citep{Winn15}. Given that a system of multiple planets undergoing convergent migration (at a rate slower than the rate of resonant libration) is expected to lock into MMR \citep{Borderies84}, the observed lack of resonant pairs led the community to conclude that either the planets never initially lock into resonance \citep[e.g.,][]{Goldreich14} or that most resonant pairs undergo a series of dynamical ``breakage'' \citep[e.g.,][]{Izidoro21}.
Recent analyses report the fraction of systems in resonance can be as high as $\sim$80\% among young ($\lesssim$100 Myr) planetary systems \citep{Dai24}, supporting the latter hypothesis.

In this research note, we report on over what timescale the initial condition set by disk-induced migration is erased by post-disk dynamical interaction, focusing on the time evolution of orbital period distribution $dN/dlogP$. Under type I migration in a solar nebular disk, $dN/dlogP$ is expected to feature a positive slope \citep{Lee17,Hallatt20} with a pile-up at the disk inner edge near the co-rotation radius with respect to the host star, as expected from the theory of magnetospheric truncation and disk locking \citep[e.g.,][]{Ghosh79}. Once the disk gas dissipates away, scattering between protoplanets can trigger a series of collisional mergers towards a configuration that is stable over $\gtrsim$Gyr timescales, with adjacent pairs of planets spaced apart by a few mutual Hill radii which scales with orbital distance $a$. It follows that at a population level, $dN/dlogP$ should be flat in log period for dynamically stable multi-planetary systems \citep{Lee17}, which is observed for sub-Neptunes \citep{Petigura18,Wilson2022}.

\begin{figure}
    \gridline{\fig{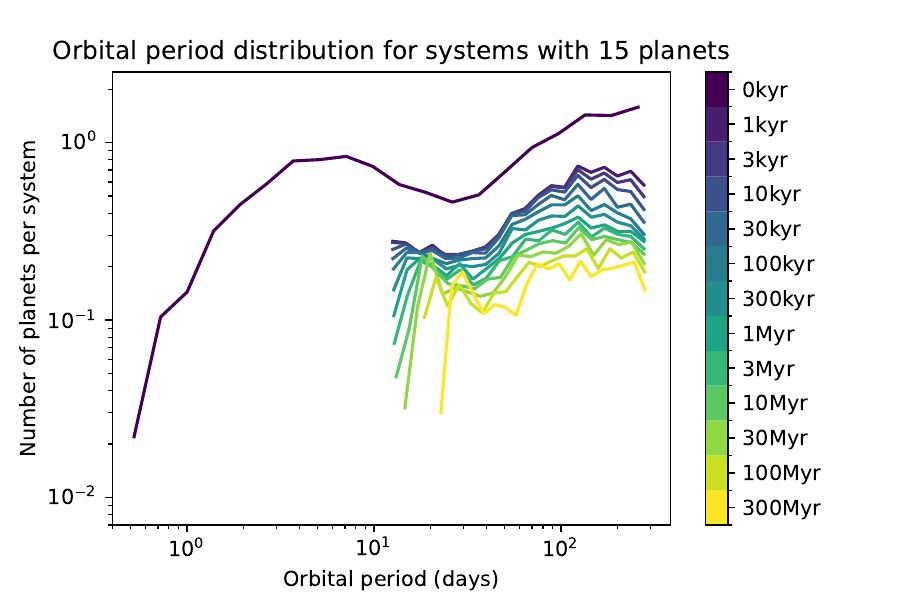}{0.5\textwidth}{}
            \fig{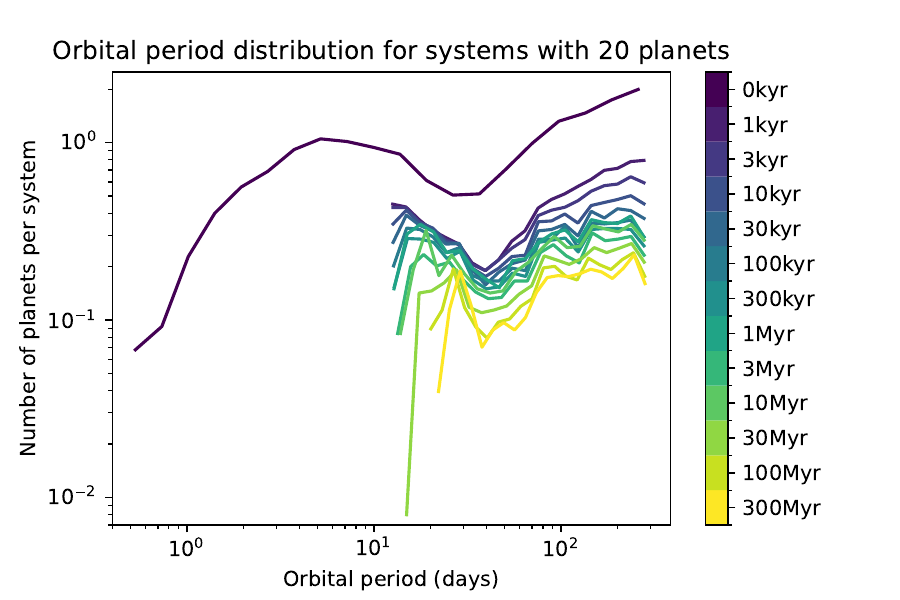}{0.5\textwidth}{}}
    \vspace{-0.5cm}
    \gridline{\fig{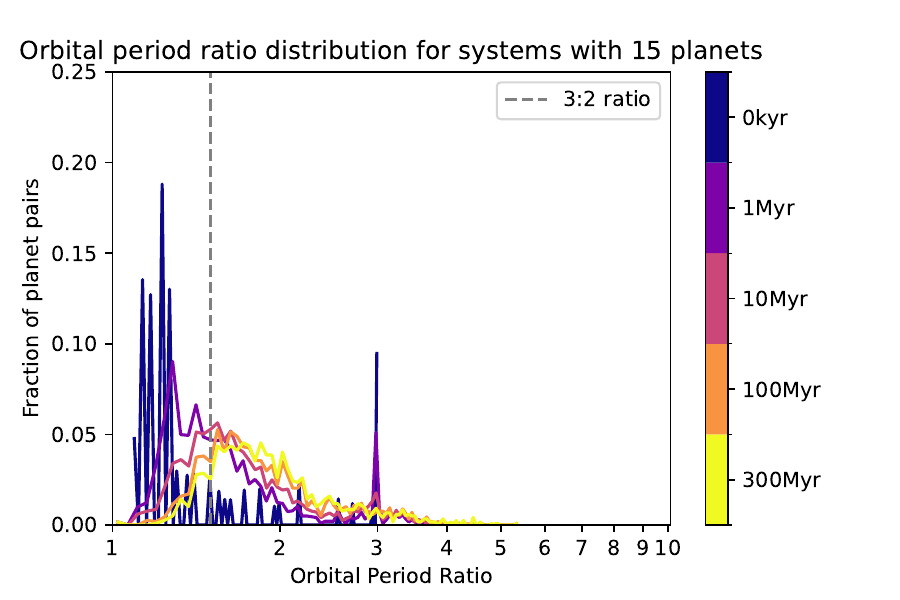}{0.5\textwidth}{}
            \fig{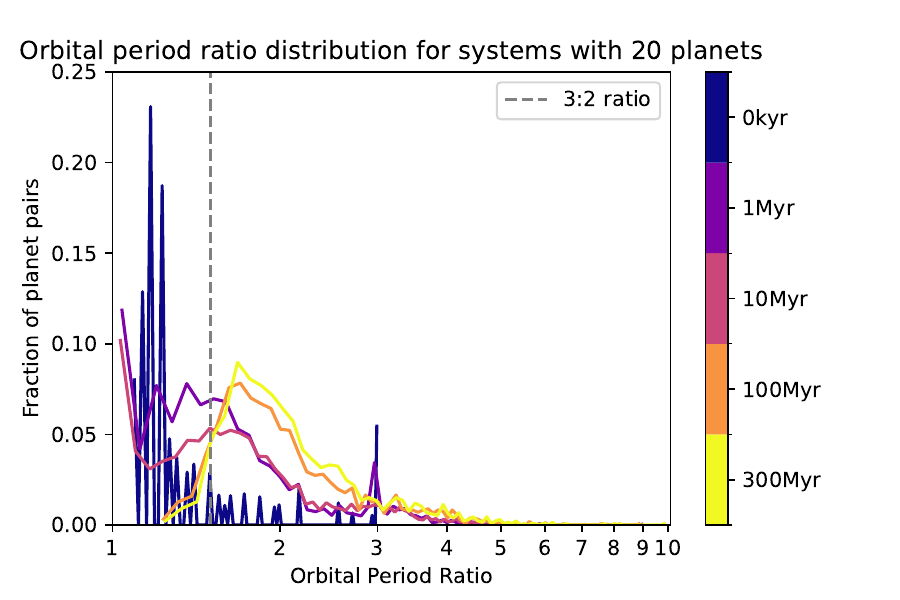}{0.5\textwidth}{}}
    \vspace{-0.5cm}
    \gridline{\fig{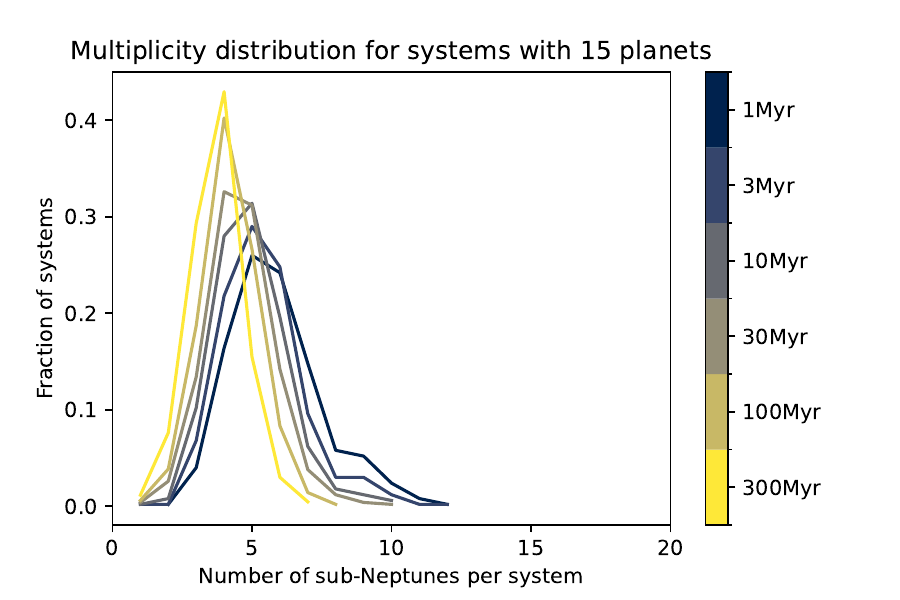}{0.5\textwidth}{}
            \fig{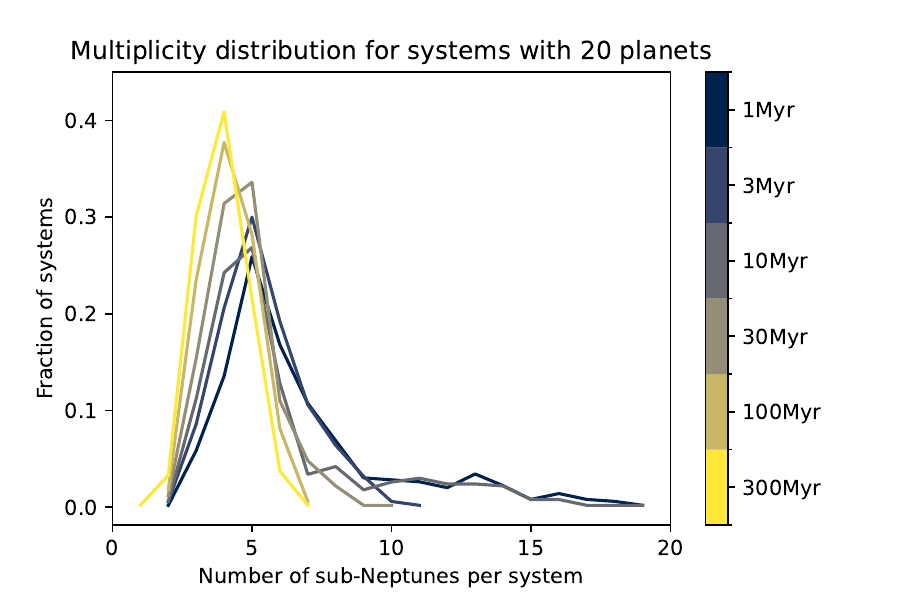}{0.5\textwidth}{}}
     \caption{Under merger events, the orbital period distribution flattens over $\sim$10-100 Myr. The immediate loss of planets inside $\sim$10 days is likely unphysical from our neglect of general relativistic precession and tidal circularization. Similarly, the period ratio distribution broadens over time reaching a broad peak near 3:2.}
    \label{fig:dNdlogP}
\end{figure}

We use REBOUND N-body integrator \citep{Rein2012} with TRACE \citep{Lu2024} to simulate the dynamical evolution of multi-planetary systems up to 300 Myr (we find similar results with the IAS15 integrator). 
We draw 2 sets of 500 randomized systems, each set containing 15 and 20 protoplanets of equal mass $M_p$=2$M_\oplus$ and radius following $R_\oplus(M_p/M_\oplus)^{1/4}$.
Planets are initially distributed according to $dN/dlogP \propto P^{0.9}$ (consistent with Type I migration in disks with gas surface density $\Sigma \propto a^{-1.28}$ and midplane temperature $T \propto a^{-3/7}$; \citealt{Hallatt20}) 
superposed with a lognormal distribution of a peak at $\sim$6 days and width $\sim$0.941, closely modeling the distribution of spin periods of pre-main sequence stars in NGC 2362 \citep{Irwin2008}. Disk-induced migration in typical solar nebula is divergent until the innermost planet parks at the inner edge. For the systems with their innermost planet placed within the inner lognormal peak, we adjust the orbital periods of subsequent planets to be at the closest peak in the orbital period ratio distribution from \citet[][their Figure 14]{Izidoro21}, leading to most planet pairs at integer ratios 7:6, 5:4, and 4:3. To ensure the distribution of orbital periods retain the expected shape, some planet pairs were adjusted to higher period ratio up to 3. For planets at orbital periods beyond 100 days, we instead shift them to a lower period ratio to achieve the same goal.
We set all initial orbits to be circular and coplanar and randomly draw the true anomalies of all planets in [0,2$\pi$]. A pair of planets is considered to have collided when they are in physical contact and we assume all collisions to lead to mergers (conserving mass, volume, and momentum), as appropriate in the inner orbits \citep[e.g.,][]{Wallace17}.

Figure \ref{fig:dNdlogP} demonstrates how post-disk dynamical scattering and merger flatten $dN/dlogP$ over time. For both 15 and 20 planet cases, the flattening becomes apparent by a few 10s to 100 Myr, suggesting that this is the timescale over which the initial condition of migration engraved in the orbital period distribution is dynamically erased. Such a timescale is roughly consistent with the finding of \citet{Dai24} who report a significant drop in the fraction of resonant pairs at $\gtrsim$100 Myr. The timescale of disruption in orbital period ratios, however, appears shorter: by $\sim$1 Myr, we see a significantly broader distribution of orbital period ratios for adjacent planet pairs. At $\sim$100--300 Myr, we still observe a noticeable peak at 3:2 period ratio, the most prevalent resonance for the observed $\lesssim$100 Myr systems \citep{Dai24}, so it could be that if our model systems are evolved beyond $\sim$1 Gyr, we may find even further disruption of resonant chains and much broader distribution of orbital period ratios, potentially in agreement with the observed drop in the resonant fractions beyond $\sim$100 Myr. We do not comment on the fine structures of resonant pairs as that is not the focus of this note \citep[see e.g.,][for a detailed investigation]{Wu24}. The distribution of planet multiplicity shows that merger events operate over a short timescale. About $\sim$30\% of systems have undergone a system-wide pairwise merger (i.e., every pair merged) by 1 Myr, and by 100 Myr, a significant fraction of systems undergo another system-wide pairwise merger ending up with a typical multiplicity of 4.

There are caveats in our calculations that need to be addressed in future work. First, Figure \ref{fig:dNdlogP} shows all planets inside of 10 days are destroyed on a very short timescale ($\lesssim$1 kyr) which we find to be due to outer planets torquing the innermost planet to fall onto the central star. We expect such behavior to become more muted when the general relativistic precession and tidal circularization are included. 
Second, our simultaneous accounting of the initial orbital period and period ratio distributions expected from disk-induced migration leads to a spurious peak at 3:1 especially for 15 planet runs. A better capture of the architectural state of planetary systems just after disk dissipation is warranted.
Finally, and related to our second caveat, we have not directly simulated the initial migration and locking into resonance while the disk gas is still around. More careful consideration of how disk-planet and planet-planet interaction jointly sculpt the initial architectural condition of multi-planetary systems is a subject of our future work.

\bibliography{rnaas}{}
\bibliographystyle{aasjournal}

\end{document}